\documentclass[twocolumn, prl,floatfix,superscriptaddress, longbibliography]{revtex4}

\usepackage{graphicx, subfigure, amsfonts,amssymb,amsmath, array, gensymb,fontenc,color, lipsum}
\usepackage{tikz}

\DeclareMathOperator{\Tr}{Tr}

\begin{document}

\title{{Quantum Electronic Transport Across ``Bite" Defects in  Graphene Nanoribbons}}
\author{Michele Pizzochero}
\email{michele.pizzochero@epfl.ch}
\author{Kristi\={a}ns \v{C}er\c{n}evi\v{c}s}
\affiliation{Institute of Physics, Ecole Polytechnique F\'ed\'erale de Lausanne (EPFL), CH-1015 Lausanne, Switzerland}
\affiliation{National Centre for Computational Design and Discovery of Novel Materials MARVEL,  Ecole Polytechnique F\'ed\'erale de Lausanne (EPFL), CH-1015 Lausanne, Switzerland}
\author{Gabriela Borin Barin}
\affiliation{Nanotech@Surfaces Laboratory, Swiss Federal Laboratories for Materials Science and Technology (EMPA), CH-8600 D\"ubendorf, Switzerland}
\author{Shiyong Wang}
\affiliation{Nanotech@Surfaces Laboratory, Swiss Federal Laboratories for Materials Science and Technology (EMPA), CH-8600 D\"ubendorf, Switzerland}
\author{Pascal Ruffieux}
\affiliation{Nanotech@Surfaces Laboratory, Swiss Federal Laboratories for Materials Science and Technology (EMPA), CH-8600 D\"ubendorf, Switzerland}
\author{Roman Fasel}
\affiliation{Nanotech@Surfaces Laboratory, Swiss Federal Laboratories for Materials Science and Technology (EMPA), CH-8600 D\"ubendorf, Switzerland}
\affiliation{Department of Chemistry and Biochemistry, University of Bern, CH-3012 Bern, Switzerland}
\author{Oleg V.\ Yazyev}
\email{oleg.yazyev@epfl.ch}
\affiliation{Institute of Physics, Ecole Polytechnique F\'ed\'erale de Lausanne (EPFL), CH-1015 Lausanne, Switzerland}
\affiliation{National Centre for Computational Design and Discovery of Novel Materials MARVEL,  Ecole Polytechnique F\'ed\'erale de Lausanne (EPFL), CH-1015 Lausanne, Switzerland}

\date{\today}

\begin{abstract}
On-surface synthesis has recently emerged as an effective route towards the atomically precise fabrication of  graphene nanoribbons of controlled topologies and widths. However, whether and to which degree structural disorder occurs in the resulting samples is a crucial issue for prospective applications that remains to be explored. Here, we experimentally identify missing benzene rings at the edges, which we name ``bite" defects, as the most abundant type of disorder in armchair nanoribbons synthesized by the bottom-up approach. First, we address their density and spatial distribution on the basis of scanning tunnelling microscopy and find that they exhibit a strong tendency to aggregate. Next, we explore their effect on the quantum charge transport from first-principles calculations, revealing that such imperfections substantially disrupt the conduction properties at the band edges. Finally, we generalize our theoretical findings to wider nanoribbons in a systematic manner, hence establishing practical guidelines to minimize the detrimental role of such defects on the charge transport. Overall, our work portrays a detailed picture of ``bite" defects in bottom-up armchair graphene nanoribbons and assesses their effect on the performance of carbon-based nanoelectronic devices.
\end{abstract}
\maketitle

\paragraph{Introduction. } 
The first isolation of graphene over a decade ago sparked a number of new avenues for designing next-generation devices in the ultimate limit of miniaturization \cite{Novo04}. Being a highly flexible, mechanically robust two-dimensional sheet featuring massless Dirac fermions and ambipolar ballistic transport of charge carriers over microscopic length scales \cite{Meu16, Novo05a,Neto09}, graphene has qualified as a promising candidate for prospective applications in novel carbon-based electronics \cite{Ares07}. Yet, the success of graphene-based logic devices has been severely hampered by the lack of a band gap in its low-energy electronic spectrum \cite{Novo04, Neto09}, hence hindering the realization of sizable on-off current ratios.

Within the wealth of strategies that have been devised to promote a band-gap opening in graphene \cite{Balo10a, Ni08a, Zhan09a, Achilli14}, quantum confinement of charge carriers -- resulting in one-dimensional graphene nanoribbons (GNRs) -- is arguably the most effective one \cite{Son06a, Yaz13}.  Conventionally, GNRs have been  realized in a ``top-down" fashion by either nanolithographing graphene flakes or unzipping nanotubes \cite{Han07, Jiao09}, yielding either irregular nanoribbons with disordered edges or providing little control over their width and chirality. 
The field has experienced a new renaissance with the first compelling demonstration that a ``bottom-up" approach relying on synthetic polymerization and subsequent cyclodehydrogenation of precursor molecules deposited on metallic surfaces \cite{Cai10a}. Since then, by carefully selecting the precursor monomers, a large number of diverse nanoribbons with desired topologies, edge shapes and widths has been manufactured \cite{Yano20, Liu15a, Vo14}, thereby achieving a large degree of tunability over their electronic properties \cite{Chen15a, Yaz13, Son06a, Llinas2017, Groning2018}. Remarkably, such a bottom-up approach has ignited the opportunity of assembling more complex two- \cite{Blan12a, Ma19a} and multi-terminal \cite{Cai10a} junctions as well as heterojunctions \cite{Bron18a, Nguy17, Cai14a}  of graphene nanoribbons, potentially paving a way towards all-carbon nanocircuitry \cite{Jaco17, Kang13, Martini19, Sun2020, ElAbbassi2020}.

Owing to their mechanical robustness, long-term stability under ambient conditions, easy transferability onto target substrates \cite{BorinBarin2019}, fabrication scalability \cite{DiGiovannantonio2018}, and suitable band-gap width \cite{Talirz17}, 9-atom wide armchair graphene nanoribbons (9-AGNRs) have emerged as the most promising candidates to be integrated as active channels in field-effect transistors (FETs). In particular, among the graphene-based electronic devices realized so far, 9-AGNR-FETs are those displaying the highest performance, with 1 $\mu$A on-currents and 10${^5}$ on-off current ratio at $V\textsubscript{D} = -1$ V \cite{Llinas2017}. 
Although the detrimental role of defects on electronic devices is well known, the performance of current GNR-FETs is limited by significant Schottky barriers at the contacts, which have prevented an experimental characterization of  the impact of GNR edge defects on the device performance. In fact, to which extent structural disorder is present in atomically precise GNRs is an issue which has not been settled to date, despite its crucial consequence on the resulting devices. 

Here, we combine experimental and theoretical efforts to investigate defects in bottom-up armchair graphene nanoribbons, with a special focus on 9-AGNRs. By means of scanning-tunnelling and atomic-force microscopies, we identify missing benzene rings at the edges as abundant defects, and additionally underly their effect on the charge transport on the basis of extensive first-principles calculations. Overall, our work offers an unprecedented view on the nature of the structural disorder in synthetic armchair graphene nanoribbons, which is instrumental to the realization of novel carbon-based electronic devices.

 \begin{figure}[]
  \centering
 \includegraphics[width=1\columnwidth]{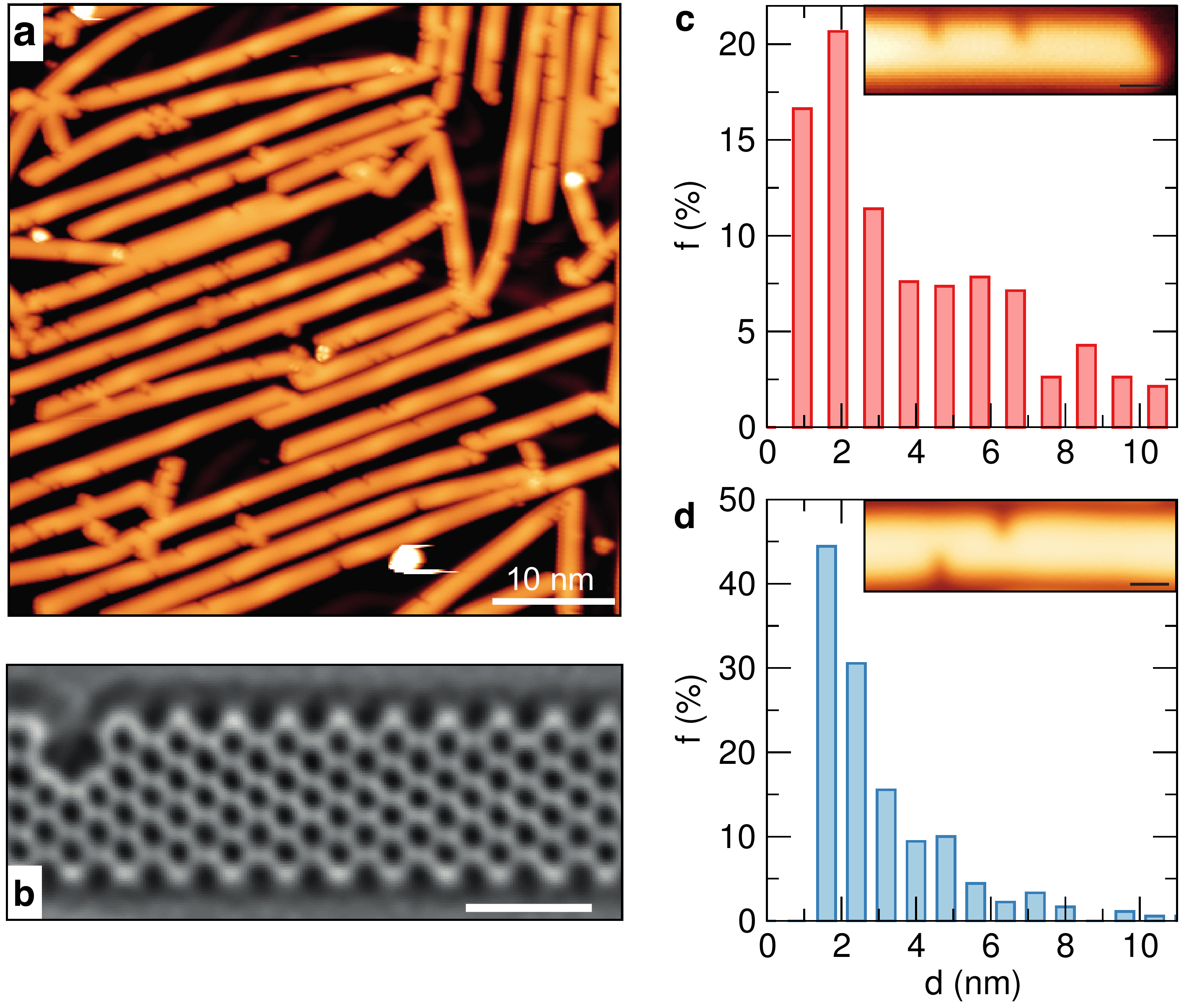}
  \caption{STM and NC-AFM characterization of 9-AGNRs on Au(111). (a) STM topography image of 9-AGNRs  (LHe, $-$1.5 V, 0.01 nA). Notice the ubiquitous presence of missing atoms at the edges. (b) Laplace-filtered NC-AFM image acquired with a CO-functionalized tip (0.01 V, 0.002 nA) of a ``bite" defect in 9-AGNR. Scale bar is 1 nm. (c,d) Positional correlation analysis of ``bite" defects in 9-AGNRs, \emph{i.e.}\ normalized frequency of occurrence ($f$) of relative distance between ``bite" defects ($d$) forming at the same edge (c) or opposite edges (d). Insets show STM images of representative configuration of defects pair. \label{Fig0}}
\end{figure}

\bigskip
\paragraph{Results and Discussion. } We synthetize 9-AGNRs by relying on the coupling and subsequent cyclodehydrogenation of 3',6'-di-iodine-1,1':2',1''-terphenyl (DITP) precursor molecule on the Au (111) surface \cite{DiGiovannantonio2018}, as we further detail in the Methods section. A representative scanning tunnelling microscopy (STM) image overviewing the resulting products is shown in Fig.\ \ref{Fig0}(a). It can be clearly observed that 9-AGNRs invariably exhibit missing atoms at the edges as a predominant type of disorder. We inspect the atomic structure of such defects through non-contact atomic-force microscopy (NC-AFM) imaging, as displayed in Fig.\ \ref{Fig0}(b). Our result reveals that these defects -- that we dub ``bite" defects -- consist of a missing benzene ring, and we estimate their density to 0.19 $\pm$ 0.10 nm$^{-1}$.  Such ``bite" defects originate from the C-C bond scission that occurs during the cyclodehydrogenation step of the reaction \cite{Talirz17}, as we show in Supplementary Fig.\ S1. Similar defects have also been observed in chevron-edged GNRs synthesized from 6,11-dibromo-1,2,3,4-tetraphenyltriphenylene precursors, and their formation was correlated with the cyclization of multiple flexible phenyl rings against each other \cite{Cai10a}. 

Furthermore, we study the spatial distribution of ``bite" defects. In Fig.\ \ref{Fig0}(c) and (d), we give the pair distribution function (\emph{i.e.}, the normalized frequency of occurrence over a wide interval of relative distances) of ``bite" defects forming either  at the same edge [Fig.\ \ref{Fig0}(c)] or at opposite edges [Fig.\ \ref{Fig0}(d)]. A representative STM image of each of the two configurations is also shown in the corresponding insets. Our analysis demonstrates that ``bite" defects strongly tend to agglomerate within approximately 2 nm, irrespective of whether the same or opposite edges are considered.  Additionally, we found that the number of defects forming at the same edge is double than that forming at opposite edges, indicating that an edge selectivity is operative. Overall, we observe that ``bite" defects preferentially form close to each other at the same edge of 9-AGNRs.

 \begin{figure}[]
  \centering
 \includegraphics[width=1\columnwidth]{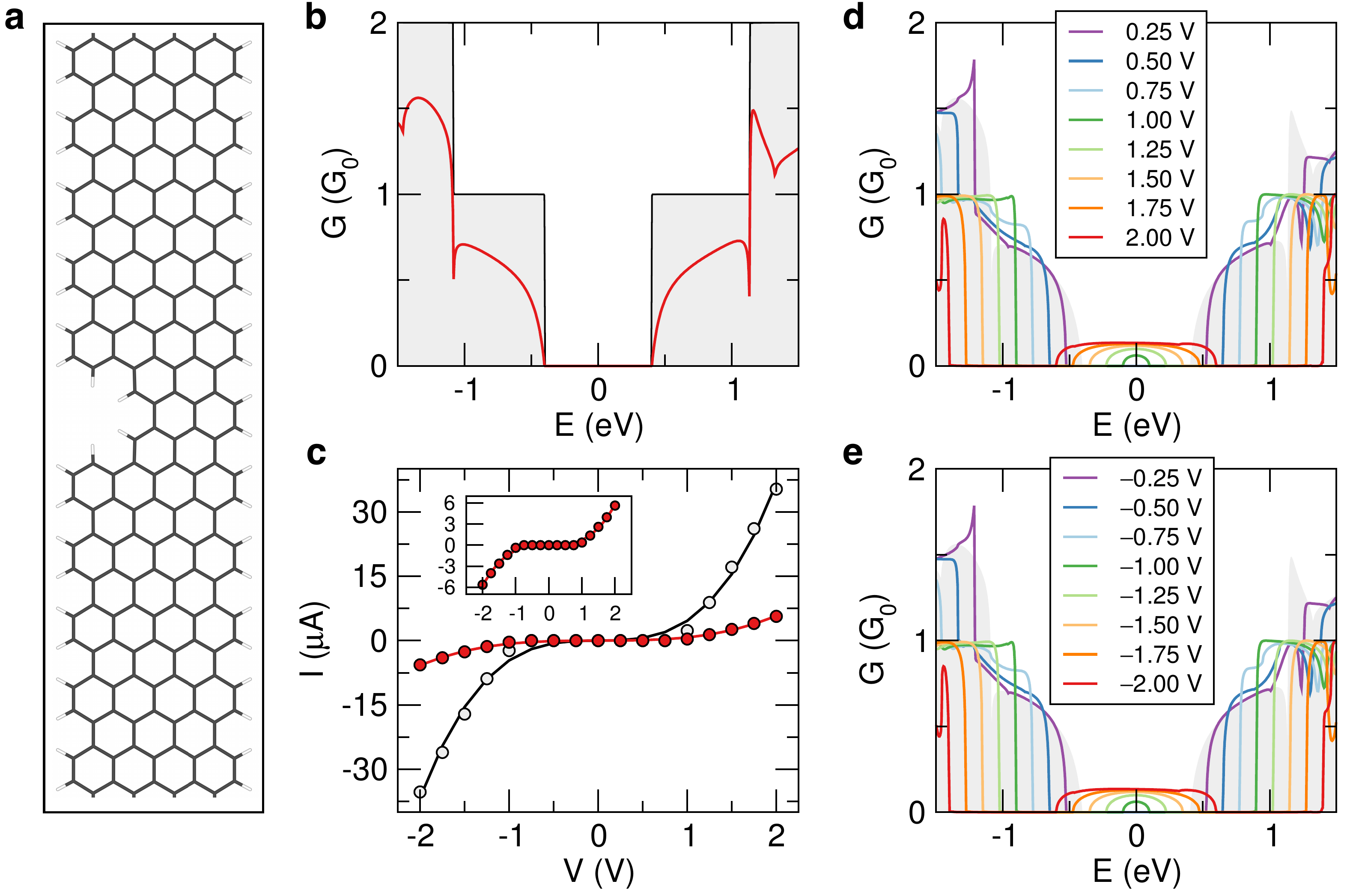}
  \caption{(a) Atomic structure of a ``bite" defect in 9-AGNR. (b) Zero-bias conductance spectra of pristine (grey) and defective (red) 9-AGNR, the latter hosting a ``bite" defect. (c) $I$-$V$ characteristics of pristine (grey) and defective (red) 9-AGNR. Circles indicate results of first-principles calculations while solid lines are the fit to the Simmon's formula in to the intermediate-voltage regime, $I \propto (V + V^3)$. Evolution of the conductance in defective 9-AGNR under (d) positive and (e) negative bias voltages with respect to the zero-bias conductance (grey), with 0.25 eV $\leq |V| \leq$ 2.00 eV. \label{Fig1}}
\end{figure}

With this systematic experimental exploration of ``bite" defects in armchair graphene nanoribbons at hand, we next address their effect on the quantum charge transport by combining density-functional theory calculations with non-equilibrium Green's function technique (see the Methods section).  We start considering a single defect in 9-AGNR, the atomic structure of which is displayed in Fig.\ \ref{Fig1}(a). In Fig.\ \ref{Fig1}(b), we show the zero-bias conductance spectrum of a defective 9-AGNR, and additionally compare our result with that of the pristine nanoribbon that shows ideal conductance quantization. It is found that the introduction of a ``bite" defect leads to a pronounced decrease of the transmission at the band edges. The transmission profiles are very similar at the edges of both the valence and conduction bands, but the electron-hole asymmetry becomes more pronounced at higher energies. In order to quantify the conductance suppression effect, we introduce a descriptor, $\tau$, which estimates the fraction of the conductance which is preserved in the vicinity ($\delta E$) of the valence band maximum (VBM) and conduction band minimum (CBM) upon the defect formation as
\begin{equation}
\tau = \int_{\textnormal{VBM}-\delta E}^{\textnormal{VBM}} \frac{G\textsubscript{d}(E)}{G\textsubscript{p}(E)} dE + \int_{\textnormal{CBM}}^{\textnormal{CBM} +\delta E}  \frac{G\textsubscript{d}(E)}{G\textsubscript{p}(E)} dE,
\label{Eqn1}
\end{equation} 
with $G\textsubscript{p}(E)$ and $G\textsubscript{d}(E)$ being the conductance of the defective and pristine armchair graphene nanoribbons, respectively. Here and below, we assume $\delta E = 0.10$ eV and find $\tau = 26$\%, thus indicating a considerable reduction of the conductance due to the presence of a ``bite'' defect at the edge of 9-AGNRs.

We extend our investigation through the determination of the charge transport properties under finite bias voltages. Fig.\ \ref{Fig1}(c) compares the $I$-$V$ characteristics of a 9-AGNR with and without a ``bite" defect. In both cases, currents arise when the applied bias voltage exceeds in magnitude the width of the band-gap ($\sim$1 eV at the adopted level of theory). Within this regime, zero-energy contributions emerge in the conductance spectra shown in Figs.\ \ref{Fig1}(d) and (e), hence signalling the enhancement of the tunnelling probability induced by the increase of the bias voltage. The current grows with the applied bias voltage following a nearly cubic scaling, as supported by the excellent agreement between the results of our calculations and the fit to the Simmon's formula appropriate to the intermediate-voltage range $I \propto (V + V^3)$ \cite{Simmons1, Simmons2}, as displayed in Fig.\ \ref{Fig1}(c).  The sign of the applied bias voltage is found to be irrelevant for shaping the evolution of both the current [Fig.\ \ref{Fig1}(c)] and conductance [Figs.\ \ref{Fig1}(d) and (e)], as a consequence of the electron-hole symmetry which is largely retained in a wide energy interval in the defective lattice. The main difference in the $I$-$V$ characteristics of pristine and defective 9-AGNR traces back to the current intensities, which are found to be lowered by one order of magnitude upon the formation of a ``bite" defect. Overall, our findings clearly pinpoint the detrimental role that ``bite" defects play on the electronic transport properties of 9-AGNRs.

 \begin{figure}[]
  \centering
 \includegraphics[width=0.85\columnwidth]{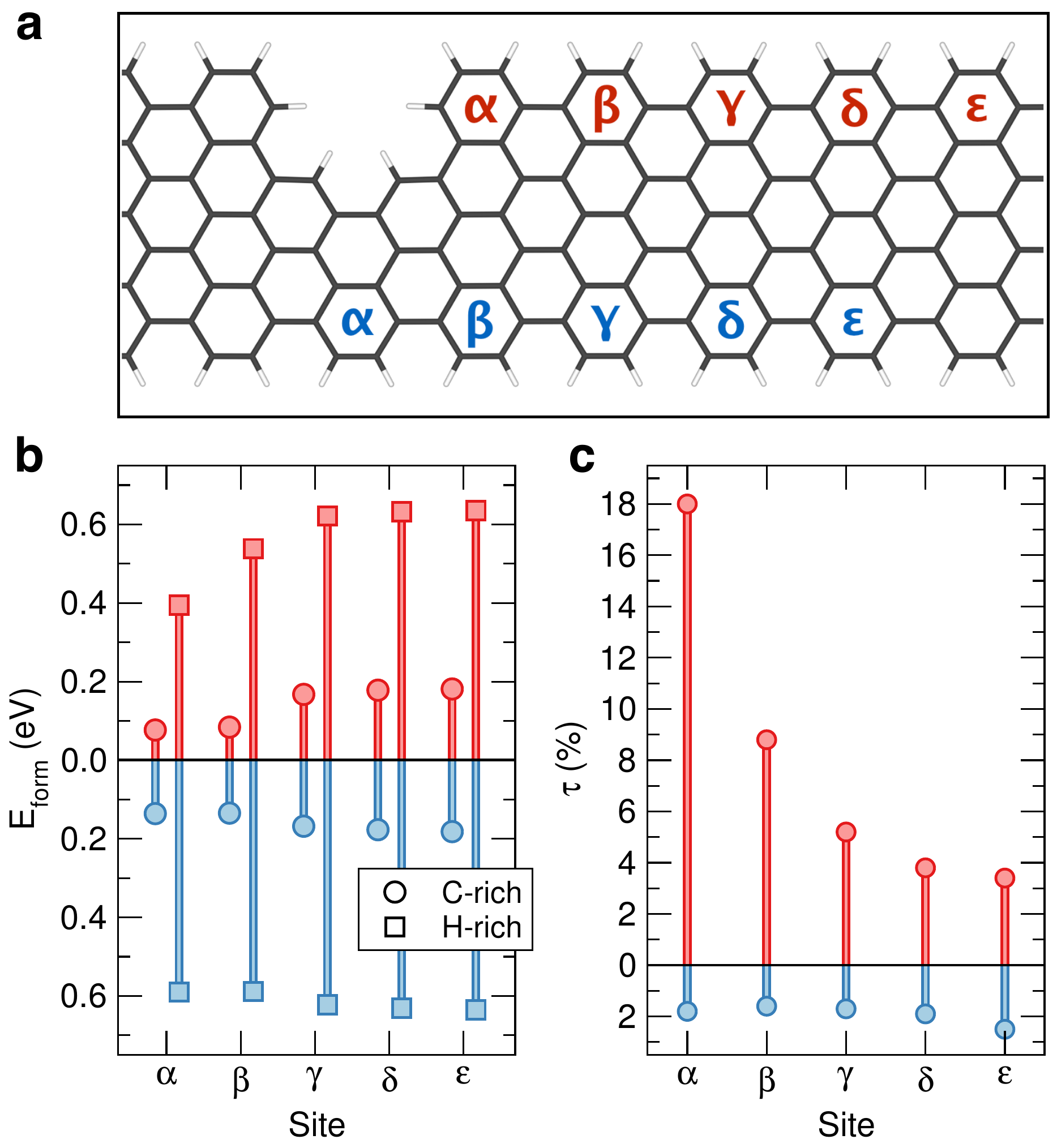}
  \caption{(a) Atomic structure of a single ``bite" defect in 9-AGNR, along with the lattice sites at which the introduction of a second ``bite" defect is introduced. (b) Formation energies of a pair of ``bite" defects forming at the sites given in panel (a) under C-rich and H-rich conditions, with defects forming at the same (opposite) edge(s) given in red (blue). (c) Evolution of the $\tau$ descriptor [given in Eqn.\ (\ref{Eqn1})] for pair of ``bite" defects forming at the lattice sites given in panel (a), either at the same edge (red) or opposite edges (blue). \label{Fig2} }
\end{figure}

\begin{figure*}[]
  \centering
 \includegraphics[width=1.75\columnwidth]{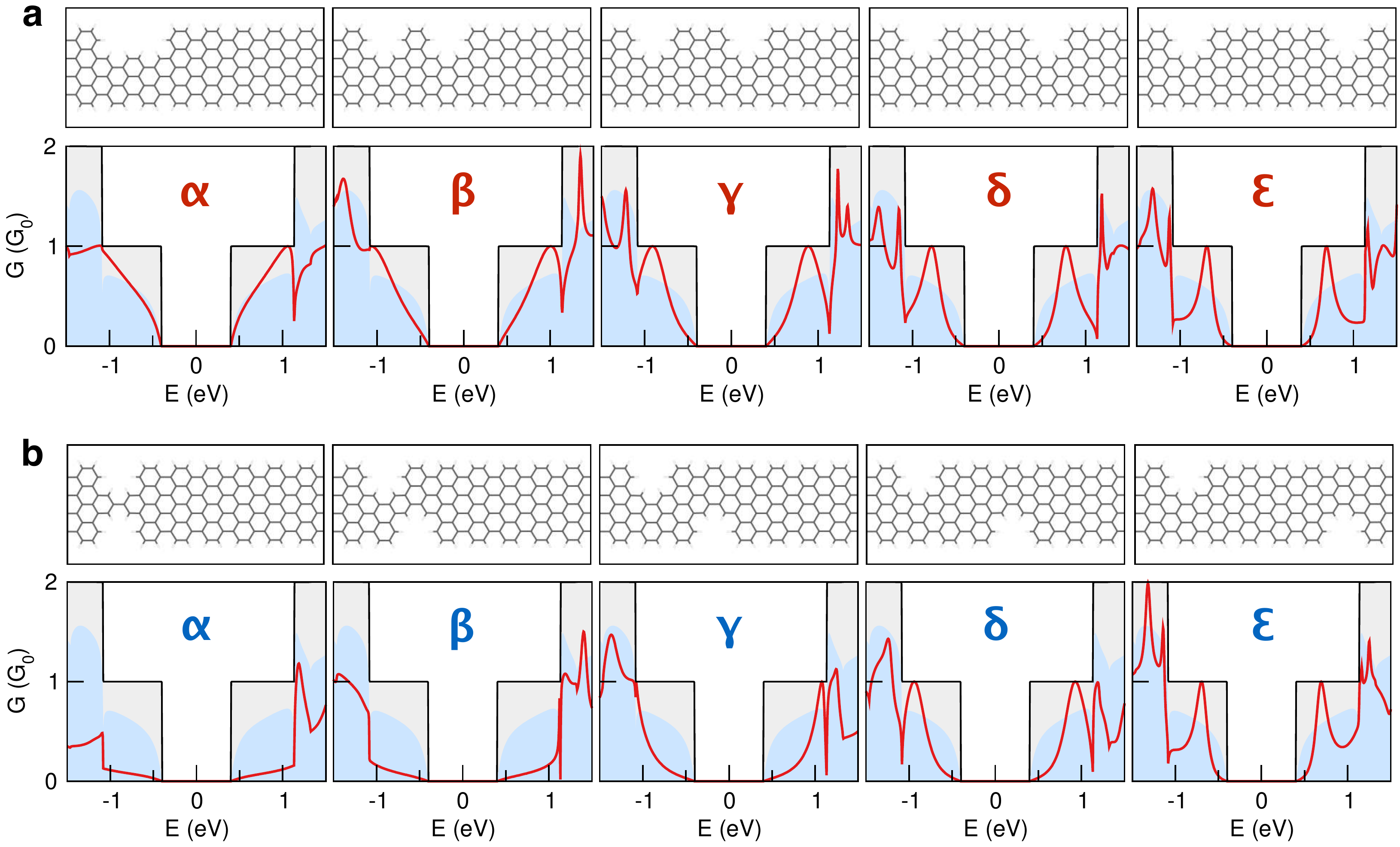}
  \caption{Conductance spectra of pairs of ``bite" defects (red) forming at (a) the same edge and (b) opposite edges, labeled according to Fig.\ \ref{Fig2}(a). Upper panels show the atomic structure of the defects configuration considered in each case. Also given for comparative purposes are the conductance spectra of 9-AGNR either in the pristine case (grey) and hosting a single ``bite" defect (light blue). \label{Fig3}}
\end{figure*}

We then address the formation of pairs of ``bite" defects, either at the same edge or at the opposite edges of the nanoribbon, of the kind shown in the insets of Fig.\ \ref{Fig0}(c) and (d). The configurations considered are presented in Fig.\ \ref{Fig2}(a), and consist in introducing a second defect at increasing distances (up to 1 nm) from the first defective site. We assess the relative stability of ``bite" defects through the determination of their formation energy $E\textsubscript{form}$, which is the primary quantity of interest when thermodynamic equilibrium prevails. As graphene nanoribbons are binary compounds, the introduction of defects changes the nominal stoichiometry, thus rendering $E\textsubscript{form}$ a linear function of the chemical potential $\mu$ of the constituent elements. The formation energy reads
\begin{equation}
E\textsubscript{form} (\mu) = E\textsubscript{d} - E\textsubscript{p} - n\textsubscript{H} \mu\textsubscript{H} +  n\textsubscript{C} \mu_\textsubscript{C},
\label{Eq1}
\end{equation}
with $E\textsubscript{d}$ and $E\textsubscript{p}$ being the total energies of the defective and pristine models, respectively, n\textsubscript{H} (n\textsubscript{C}) the number of added (removed) H (C) atoms required to create the defect, while $\mu_\textsubscript{C}$  and  $\mu_\textsubscript{H}$ are the corresponding chemical potentials. As usual, we assume that C and H are in thermal equilibrium with armchair graphene nanoribbons of general formula C\textsubscript{x}H\textsubscript{y} through the equality $\mu\textsubscript{C\textsubscript{x}H\textsubscript{y}} = x\mu\textsubscript{C} + y \mu\textsubscript{H}$, where graphene and molecular hydrogen are assumed to be the reference systems for the determination of the chemical potential reservoirs. In Fig.\ \ref{Fig2}(b), we present the formation energy of pairs of ``bite" defects. Our calculations indicate that their stability is enhanced when two defects are in proximity to each other (that is, sites $\alpha$ and $\beta$), with the formation energy attaining the lowest value when the second defect forms at nearest-neighboring site ($\alpha$) at the same edge.  In contrast, when the distance between the two defects is larger than $\sim$9 {\AA} (\emph{i.e.}, sites $\gamma$, $\delta$, $\epsilon$),  $E\textsubscript{form}$  reaches its maximum and remains practically constant. Comparison of the results in Fig.\ \ref{Fig2}(b) with twice the formation energy of a single ``bite" defect (0.64~eV and 0.16~eV in H-rich and C-rich conditions, respectively), suggests that, under thermodynamic equilibrium, ``bite" defects exhibit a tendency to aggregate at shortest distances (positions $\alpha$ and $\beta$), but the interaction is negligible at larger distances. From a qualitative point of view, this theoretical result parallels the experimental analysis of Fig.\ \ref{Fig0}, although formation energies reflect the stability of defects under equilibrium, whereas the synthesis of 9-AGNRs (and of the defects emerging therein) is largely governed by the kinetic control.

In Fig.\ \ref{Fig3}, we overview the conductance spectra of pairs of ``bite" defects in each of the ten configurations considered. In all cases, the conductance in the vicinity of the band edges is further reduced as compared to that of the nanoribbon containing either a single or no ``bite" defect. However, the degree to which this reduction occurs is largely controlled by the relative position of the defects. This can be clearly observed in Fig.\  \ref{Fig2}(c), in which we report the values of the descriptor $\tau$ given in Eqn.\ (\ref{Eqn1}) for pairs of ``bite" defects. Depending on whether the defects are introduced at the same edge or at opposite edges, two distinct situations are identified. On the one hand, for defects forming at the same edges, $\tau$ decreases as the distance between defects increases. Specifically, the formation of a second defect at the $\alpha$ site preserves the largest amount of conductance ($\tau$ = 18\%, only slightly lower than the value obtained in the single ``bite" defect case discussed above, $\tau$ = 26\%). It is worth noticing in this context that this configuration is the thermodynamically stable one, see Fig.\  \ref{Fig2}(b). On the other hand, the formation of a pair of defects at opposite edges yields an almost complete suppression of the conductance in the vicinity of the band edges ($\tau$ $\approx$ 2\%), with $\tau$ being insensitive to the specific defects configuration. Additional results concerning the stability and conductance of pairs of ``bite" defects at opposite edges of AGNR are given in the Supplementary Fig.\ S2.

  \begin{figure*}[]
  \centering
 \includegraphics[width=1.75\columnwidth]{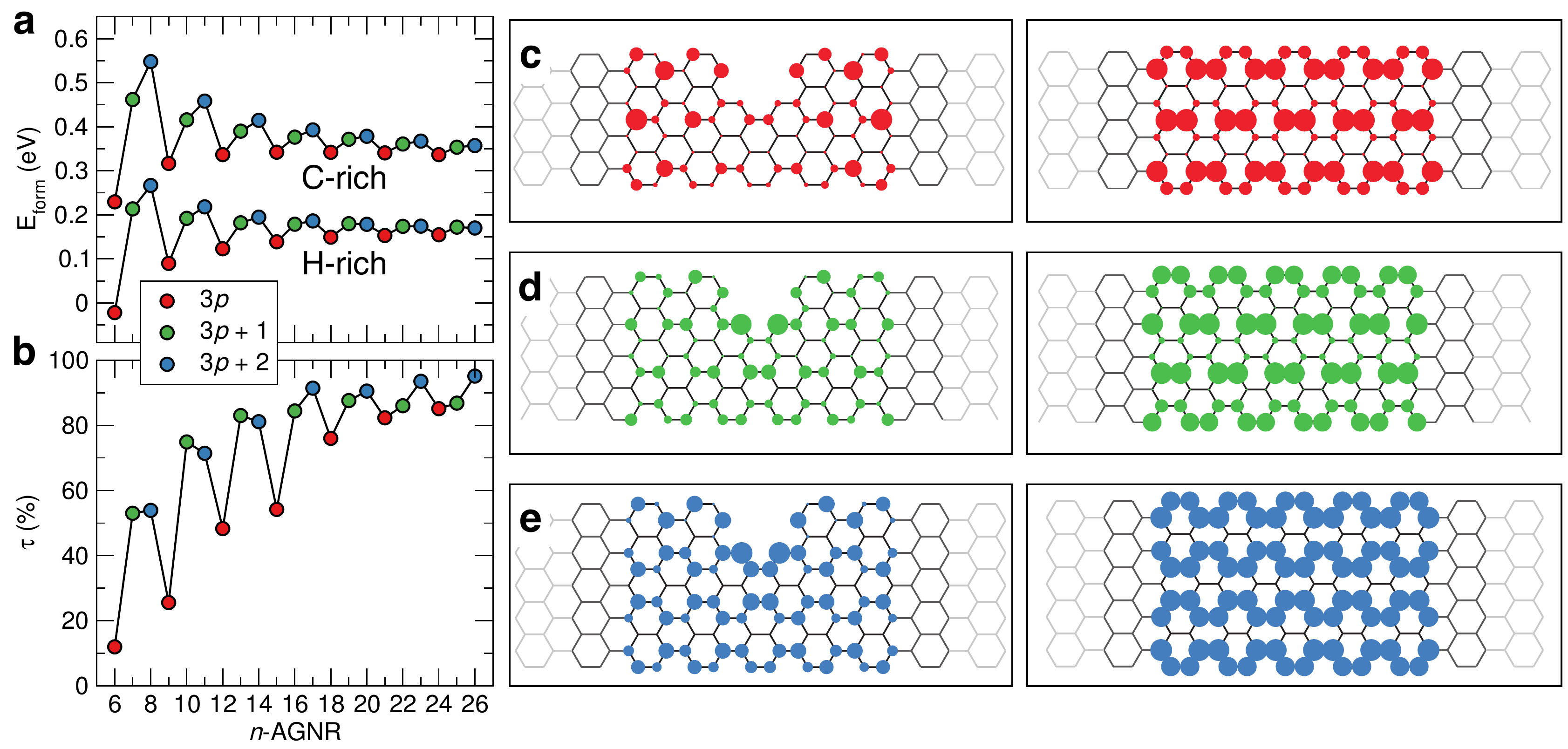}
  \caption{(a) Formation energy  and (b) $\tau$ of a single ``bite" defect in armchair graphene nanoribbons as a function of the increasing width $n$. (b) Local density of states 0.05 eV below the valence band edge of defective (left panels) and pristine (right panels) armchair nanoribbons in (c) 9-, (d) 10-, and (e) 11-AGNR, as obtained at the tight-binding level of theory.
  \label{Fig4}}
\end{figure*}

We next generalize our findings by comprehensively investigating the dependence of  the stability and charge transport of ``bite" defects on the width of the armchair graphene nanoribbon. We consider a range of widths ($n$) spanning an interval from 6 up to 26 atoms. The calculated properties show clear multiple-of-3 oscillations (Fig.\ \ref{Fig4}(a),(b)), therefore we group the nanoribbons into the $n$ = 3$p$, 3$p$+1 and 3$p$+2 families according to their width, where $p$ is a positive integer (2 $\leq p \leq$ 8 in the considered range of $n$). The stability of ``bite" defects as a function of the increasing width of the hosting armchair graphene nanoribbon is given in Fig.\ \ref{Fig4}(a). The formation energy is highly sensitive to $n$. Specifically, this value increases with $n$ in the  3$p$ family, whereas it decreases  when $n$ = 3$p$+1 or 3$p$+2. In all families, however, it remains approximately constant when the width is larger than 17 atoms. This finding suggest that, under thermodynamic equilibrium, narrow armchair nanoribbons belonging to the 3$p$ family are the most susceptible systems to ``bite" defects.

\begin{figure*}[]
  \centering
 \includegraphics[width=1.75\columnwidth]{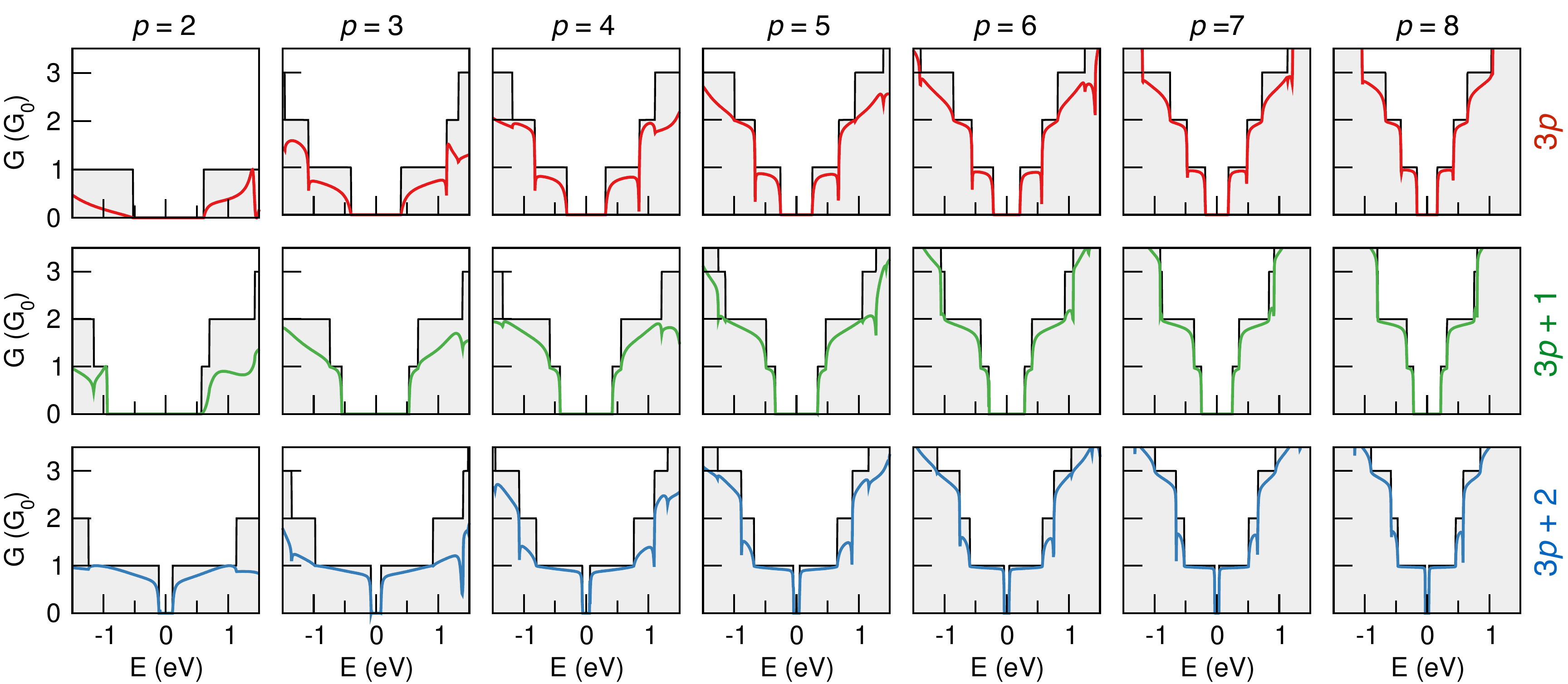}
  \caption{Transmission spectra of pristine (grey area) and defective (colored lines) armchair graphene nanoribbons for different widths, grouped according to their width into the 3$p$ (red lines),  3$p$+1 (green lines), and $3p$+2 (blue lines) families. \label{Fig5}}
\end{figure*}

Finally, we broaden our study by addressing the charge transport in defective AGNRs of varying width. The conductance spectra upon ``bite" defect formation in the  3$p$, 3$p$+1 and 3$p$+2 families are shown in Fig.\ \ref{Fig5}.  From a qualitative point of view, the disruption of the conductance is seen to be milder in wider nanoribbons as compared to the narrower ones. However, for a given value of $p$, the conductance in the vicinity of the band edges of 3$p$ armchair nanoribbons undergoes the most drastic reduction. These effects can be translated on a quantitative basis through the determination of $\tau$ as a function of $n$, as given in Fig.\ \ref{Fig4}(b).  As far as the transport properties are concerned, the conductance increases with increasing the nanoribbon width, with $\tau$ exceeding 85\% in the case of armchair graphene nanoribbons of $n$ larger than 22 atoms. This naturally reflects the decrease of edge-to-bulk ratio, end hence the effect of edge defects, upon increasing the width. Of the three families, however, we found that the conductance is the most strongly reduced in the  3$p$ one. Indeed, $\tau$ is approximately halved in the 3$p$ family as compared to both 3$p$+1 and 3$p$+2, when the same value of $p$ is considered. 

We rationalize the largest ``bite" defect-induced disruption of the conductance observed in the 3$p$ family in terms of localization effects. In Fig.\ \ref{Fig4}(c), (d), and (e) we present the local density of states calculated at the tight-binding level (see the Methods section and Supplementary Fig.\ S3) slightly below the valence band edge for both pristine and defective 9-, 10-, and 11-AGNR, as representative members of 3$p$, 3$p$+1, 3$p$+2 families, respectively (see Supplementary Fig.\ S4 for additional results). Already in the pristine armchair nanoribbons, the local density of states behave differently in the three families, being mostly localized on the two inner (outer) edge atoms in the 3$p$ $(3p+1)$ family, and rather uniformly distributed on the edges of the 3$p$+2 family. Upon the introduction of the ``bite" defect, in $3p$-AGNR we observe the emergence of a pronounced localization on the sublattice that is the closest to the defect, strongly affecting the local density of states across the entire width, and eventually leading to a significant decrease of the conductance.  In contrast, albeit a weaker localization around the ``bite" defect is observed in the $3p+1$ and $3p+2$ families as well, the local density of states far from from the defective site is substantially less altered in these systems, closely resembling that of the corresponding pristine nanoribbons.  Furthermore, we notice that such localization effects become less relevant when the width of the nanoribbon is increased (see Supplementary Fig.\ S4). Indeed, at larger values of $n$, the local density of states in the regions far from the defective site remains nearly unperturbed, thereby explaining the sizable recovery of conductance at the band edges in wide AGNRs containing a ``bite" defects, as presented in Fig.\ \ref{Fig4}(b).

\bigskip
\paragraph{Summary and Conclusions.} In summary, we have experimentally identified the ``bite" defects, \emph{i.e.}, missing benzene rings at the edges, as the predominant source of atomic-scale disorder in atomically precise 9-atom-wide armchair graphene nanoribbons. These defects form upon phenyl-ring cleavage occurring during the cyclo-dehydrogenation step of their synthesis, and exhibit a substantial tendency to aggregate within $\sim$2 nm. Our first-principles calculations reveal that ``bite" defects dramatically disrupt the charge transport of 9-AGNRs by reducing the transmission in the vicinity of the band edges from 74 \% up to  98 \%, depending on the number and configuration of defects considered. Additionally, we have expanded our theoretical investigation to $n$-AGNR (with 6 $\leq n \leq$ 26), and found that conduction properties become less sensitive to ``bite" defects in wider nanoribbons and specifically in those belonging to the 3$p$+1 and 3$p$+2 families. Altogether, we suggest that the precursor molecule or the conditions employed in the 9-AGNRs synthesis need to be re-examined in order to fabricate ``bite" defect-free nanoribbons for high-performance applications in electronic devices. Alternatively,  $n$-AGNRs with $n \neq 3p$ qualify as better candidates to minimize the impact of such structural disorder on the electronic properties. To conclude, our work uncovers the role of ``bite" defects on the charge transport of armchair graphene nanoribbons and establishes useful guidelines to mitigate their detrimental impact on the resulting electronic devices.

\section{Methods}

\paragraph{Experimental.} 9-AGNRs were synthesized from 3',6'-di-iodine-1,1':2',1''- terphenyl (DITP) as the precursor monomer under ultrahigh vacuum (UHV) conditions \cite{DiGiovannantonio2018}. First, the Au(111) (MaTeck GmbH) surface was cleaned by repeated cycles of Ar\textsuperscript{+} sputtering (1 keV) and annealing (470 \textsuperscript{o}C). Next, the precursor molecules were thermally evaporated (95 \textsuperscript{o}C) on the clean surface. Finally, the substrate was heated (0.5 \textsuperscript{o}C/s) up to 200 \textsuperscript{o}C with a 10 minute holding time to activate the polymerization reaction, followed by annealing at 400 \textsuperscript{o}C (0.5 \textsuperscript{o}C/s with a holding time of 10 minutes) to form the GNRs \emph{via} cyclodehydrogenation of the polyphenylene precursors.  A total of 445 pair of defects in the same edge and 225 pair of defects in opposite edges were analyzed using Igor Pro.

\paragraph{First-principles calculations.} First-principles calculations have been performed within the density-functional theory framework, as implemented in \textsc{siesta} \cite{SIESTA}. We treated the exchange and correlation effects under the generalized gradient approximation of Perdew, Burke, and Ernzerhof \cite{PBE}. Core electrons were described by separable norm-conserving pseudopotentials \cite{PSEUDO}, while the Kohn-Sham wavefuctions of valence electrons were expanded in a linear combination of atomic orbitals of double-$\zeta$ polarization (DZP) quality. Real space integrations have been performed with a 450 Ry mesh cutoff. The Brillouin zone was sampled with the equivalent of 21 $\times$ 1 $\times$ 1 $k$-mesh per unit cell in all cases but transport calculations, for which it was increased to  400 $\times$ 1 $\times$ 1. We optimized the atomic coordinates until the residual force acting on each atom converges to 0.01 eV/{\AA}. We introduced single ``bite" defects in otherwise pristine 7 $\times$ 1 $\times$ 1 supercells of $n$-AGNR of increasing widths (6  $\leq n \leq$ 26), containing from 112 (6-AGNR) to 392 atoms (26-AGNR). Pairs of ``bite" defects in 9-AGNR are modelled in a 14 $\times$ 1 $\times$ 1 supercell containing 308 atoms. Replicas along non-periodic directions are separated by a vacuum region larger than 10 {\AA}.

\paragraph{Tight-binding calculations.} We relied on a tight-binding model Hamiltonian with one $p_z$ orbital per carbon atom with the help of \textsc{kwant} package \cite{KWANT}. The Hamiltonian is expressed as $H = \sum_i \epsilon_i c_i^{\dagger} c_i - t \sum_{i,j} (c_i^{\dagger} c_i + H.c.)$, where $\epsilon_i$ is the on-site energy acting on the $i$-t site, $t$ the nearest-neighbor hopping integral,  $c^{\dagger}$ ($c$) the operator that creates (annihilates) an electron on site $i$ \cite{Neto09}. As proposed in Ref.\ \cite{Hancock}, we included first, second, and third nearest-neighbor hopping integrals with corresponding values of --2.70 eV, --0.20 eV, and --0.18 eV, respectively, while on-site repulsion is neglected. This tight-binding Hamiltonian yields results in excellent accord with density-functional theory calculations, as we show in the Supplementary Fig.\ S2.

\paragraph{Quantum electronic transport calculations.}
In order to investigate the quantum transport properties of armchair graphene nanoribbons, Hamiltonians obtained from density-functional theory and tight-binding calculations were next combined with the non-equilibrium Green's function formalism, as implemented in \textsc{transiesta} \cite{TRANSIESTA} and \textsc{kwant} \cite{KWANT}, respectively.  As is customary,  the transmission coefficient $T(E)$ of charge carriers of energy $E$ flowing from the source to the drain electrode is determined through the usual $T(E) = \Tr[\Gamma_R(E) G^{R}(E) \Gamma_L G^{R \dagger}(E)]$, where $\Gamma_{L, R} = i[\Sigma_{L, R} -  \Sigma^{\dagger}_{L, R}]$ is the broadening function originating from to the coupling between the source (L) and the drain (R) electrodes separated by the the central scattering region, as quantified by the self-energies  $\Sigma_{L,R}$, and $G^R = [(ES - H - \Sigma_L - \Sigma_R)]^{-1}$ is the retarded Green's function, with $H$ and $S$ being the Hamiltonian and the overlap matrix, respectively. From the obtained $T(E)$, the conductance $G(E)$ is achieved using the Landauer formula, that is $G(E) = G_0 \int T(E) -\frac{\partial f(E,T)}{\partial E} dE$, where $G_0 = \frac{2e^2}{h}$ is the conductance quantum and $f(E, T)$ is the Fermi-Dirac distribution at a given electronic temperature $T$. Finite-bias calculations have been performed in a self-consistent manner.

\section{Supporting Information}
Scheme of the chemical reaction yielding defective 9-AGNR, along with additional first-principles calculations concerning the thermodynamic stability and quantum electronic transport of pairs of ``bite" defects, assessment the performance of the adopted tight-binding models against density-functional theory results, and local density of states in pristine and defective $n$-AGNR, with 9 $\leq n \leq$ 14. 

\section{Acknowledgments} M.P.\ acknowledges QuanSheng Wu at EPFL for technical assistance. M.P., K.\v{C}.\ and O.V.Y.\ are financially supported by the Swiss National Science Foundation (Grant No.\ 172543) and NCCR MARVEL. G.B.B., P.R., and R.F. acknowledge funding by the Swiss National Science Foundation under the Grant No.\ 20PC21-155644, the European Union Horizon 2020 research and innovation program under grant agreement No.\ 785219 (Graphene Flagship Core 2), and the Office of Naval Research BRC Program under the grant N00014-12-1-1009. First-principles calculations have been performed at the Swiss National Supercomputing Center (CSCS) under the projects s832 and s1008.
 \bibliography{References}

\end{document}